\documentclass[event]{sigirforum}

\def\pubissue{Vol. 57 No. 1 -- June 2023}

\def\eventdate{18 July 2024}
\def\eventurl{https://llm4eval.github.io}

\usepackage{xcolor}
\usepackage{booktabs}

\begin{document}
\title{Report on the 1st Workshop on Large Language Model for Evaluation in Information Retrieval (\workshopname 2024) at SIGIR 2024}

\authors{
\author[\scriptsize hossein.rahmani.22@ucl.ac.uk]{Hossein A.~Rahmani}{University College London}{London, UK}
\and
\author[c.n.siro@uva.nl]{Clemencia Siro}{University of Amsterdam, }{Amsterdam, The Netherlands}
\and
\author[m.aliannejadi@uva.nl]{Mohammad Aliannejadi}{University of Amsterdam}{Amsterdam, The Netherlands}
\and
\author[nickcr@microsoft.com]{Nick Craswell}{Microsoft}{Seattle, US}
\and
\author[claclark@gmail.com]{Charles L.~A.~Clarke}{University of Waterloo}{Ontario, Canada}
\and
\author[guglielmo.faggioli@unipd.it]{Guglielmo Faggioli}{University of Padua}{Padua, Italy}
\and
\author[bmitra@microsoft.com]{Bhaskar Mitra}{Microsoft}{Montréal, Canada}
\and
\author[pathom@microsoft.com]{Paul Thomas}{Microsoft}{Adelaide, Australia}
\and
\author[emine.yilmaz@ucl.ac.uk]{Emine Yilmaz}{University College London}{London, UK}
}

\newcommand{\partitle}[1]{\vspace{2mm}\noindent\textbf{#1}}
\newcommand{\workshopname}{\texttt{LLM4Eval}\xspace}
\newcommand{\ie}{\emph{i.e.}}
\newcommand{\eg}{\emph{e.g.}}
\newcommand{\todo}[1]{\textcolor{orange}{[ToDo:] #1}}
\newcommand{\saeed}[1]{\textcolor{blue}{[Saeed:] #1}}
\newcommand{\updated}[1]{\textcolor{brown}{#1}}

\maketitle 
\begin{abstract}
The first edition of the workshop on Large Language Model for Evaluation in Information Retrieval (\workshopname 2024) took place in July 2024, co-located with the ACM SIGIR Conference 2024 in the USA (SIGIR 2024). The aim was to bring information retrieval researchers together around the topic of LLMs for evaluation in information retrieval that gathered attention with the advancement of large language models and generative AI. Given the novelty of the topic, the workshop was focused around multi-sided discussions, namely panels and poster sessions of the accepted proceedings papers.
\end{abstract}

\section{Introduction}
\label{sec:introduction}

Large language models (LLMs), such as ChatGPT\footnote{\url{https://chatgpt.com}}, have demonstrated increasing effectiveness, with larger models performing well on tasks where smaller models are insufficient. Recently, LLMs have been actively explored for various evaluation tasks, among others.

In information retrieval (IR), among other applications, LLMs are actively explored for estimating query-document relevance, both for ranking and for label generation \citep{rahmani2024synthetic,craswell2024overview}. The latter can then be used for training and evaluating other less powerful but more efficient rankers. LLMs are employed for relevance labeling in industry ~\citep{thomas2023large}. The evaluation methodologies apply a wider range of LLMs and prompts to the labeling problem, potentially addressing a broader range of quality issues.

Motivated by these observations, we believed that a workshop on evaluation strategies in the context of LLMs would question whether IR and NLP were truly facing a paradigm shift in evaluation strategies. Therefore, we organized this workshop to provide a fresh perspective on LLM-based evaluation through an information retrieval lens. The workshop also provided an opportunity to reflect on the benefits and challenges of LLM-based evaluation in academia and industry. Finally, we encouraged submissions and discussions on further evaluation topics and models, where existing literature is scarce, such as recommender systems, learning to rank, and diffusion models.

This paper is an event report of our own \workshopname \citep{rahmani2024llm4eval} event, the first workshop on Large Language Model for Evaluation in Information Retrieval (\workshopname 2024), held in conjunction with SIGIR 2024. The workshop had a poster session with accepted papers and a panel discussion. We report on how we organized the workshop (Section \ref{sec:workshop-overview}), provide a descriptive account of what happened at the workshop (Section \ref{sec:workshop-program}), and report on what we learned from the \texttt{LLMJudge} challenge (Section \ref{sec:llmjudge}).
\section{Workshop Overview}
\label{sec:workshop-overview}

This section provides a descriptive account of the paper review process and how we organized the workshop and panel session. We begin by defining what topics the workshop was mainly focused on among many others.

\subsection{Topics}
The workshop focused on models, techniques, data collections, and methodologies for information retrieval evaluation in the era of LLMs. These include but are not limited to:

\begin{itemize}
    \item LLM-based evaluation metrics for traditional IR and generative IR
    \item Agreement between human and LLM labels
    \item Effectiveness and/or efficiency of LLMs to produce robust relevance labels
    \item Investigating LLM-based relevance estimators for potential systemic biases
    \item Automated evaluation of text generation systems
    \item End-to-end evaluation of Retrieval Augmented Generation systems
    \item Trustworthiness in the world of LLMs evaluation
    \item Prompt engineering in LLMs evaluation
    \item Effectiveness and/or efficiency of LLMs as ranking models
\end{itemize}

\subsection{Format}
The workshop was a full-day in-person workshop held in Washington D.C., US on the 18th of July 2024. The day was organized as follows:

\begin{table}[ht]
  \caption{The detailed program for the \workshopname Workshop at SIGIR 2024.}
  \centering
  \label{tbl:format}
  \begin{tabular}{ll}
    \toprule
    Time & Agenda \\
    \midrule
    \multicolumn{2}{l}{Morning} \\
    \midrule
    9:00 $-$ 9:15	& Opening Remarks \\
    9:15 $-$ 10:00	& Keynote 1: \textbf{Ian Soboroff, NIST} \\
    10:00 $-$ 10:30	& Booster Talks 1 \\
    10:30 $-$ 11:00	& Coffee Break \\
    11:00 $-$ 11:30	& Booster Talks 2 \\
    11:30 $-$ 12:30	& Poster Session \\
    12:30 $-$ 13:30	& Lunch \\
    \midrule
    \multicolumn{2}{l}{Afternoon} \\
    \midrule
    13:30 $-$ 14:15	& Keynote 2: \textbf{Donald Metzler, Google} \\
    14:15 $-$ 14:30	& \texttt{LLMJudge} Presentation \\
    14:30 $-$ 15:00	& Discussion on the results of \texttt{LLMJudge} \\
    15:00 $-$ 15:30	& Coffee Break \\
    15:30 $-$ 16:55	& Panel Discussion \\
    \bottomrule 
  \end{tabular}
\end{table}

\subsection{Program Committees}
\label{sec:reviewers}
\workshopname exists thanks to the dedication of $24$ researchers who volunteered their time to review the submissions. We are deeply grateful to each member for their commitment to the workshop. Below is a list of the program committee members:

\begin{itemize}
\item Zahra Abbasiantaeb, University of Amsterdam
\item Mofetoluwa Adeyemi, University of Waterloo
\item Marwah Alaofi, RMIT University
\item Negar Arabzadeh, University of Waterloo
\item Shivangi Bithel, IIT Delhi
\item Francesco Luigi De Faveri, University of Padua
\item Yashar Deldjoo, Polytechnic University of Bari
\item Gianluca Demartini, The University of Queensland
\item Laura Dietz, University of New Hampshire
\item Yue Feng, UCL
\item Claudia Hauff, Spotify
\item Bhawesh Kumar, Verily Life Sciences
\item Yiqun Liu, Tsinghua University
\item Sean MacAvaney, University of Glasgow
\item James Mayfield, Johns Hopkins University
\item Chuan Meng, University of Amsterdam
\item Ipsita Mohanty, Carnegie Mellon University
\item Mohammadmehdi Naghiaei, University of Southern California
\item Pranoy Panda, Fujitsu Research
\item Lu Wang, Microsoft
\item Xi Wang, University of Sheffield
\item Orion Weller, Johns Hopkins University
\item Jheng-Hong Yang, University of Waterloo
\item Oleg Zendel, RMIT University
\end{itemize}
\section{Workshop Program}
\label{sec:workshop-program}
In this section, we present an overview of the \workshopname workshop, encompassing details about its participants, accepted papers, poster sessions, panel discussion.

\subsection{In Numbers}
\label{sec:numbers}
As the first workshop on LLMs for evaluation in information retrieval, \workshopname 2024 has attracted significant interest from the community. The workshop welcomed more than 50 in-person participants, reflecting the growing curiosity and engagement around the evolving role of LLMs in information retrieval evaluation.

\subsection{Keynotes}
\label{sec:keynotes}
\workshopname featured two invited keynote talks. We present the title and abstract of each talk below along with the name of each speaker.

\subsubsection{Keynote 1: A Brief History of Automatic Evaluation in IR \\ \textit{by} Ian Soboroff, NIST}
\textbf{Abstract.} The ability of large language models such as GPT4 to respond to natural language instructions with flowing, grammatical text that reflects world knowledge has generated (sorry) significant interest in IR, as it has everywhere, and specifically in the area of IR evaluation. It seems that just as we “prompt” a human assessor to provide a relevance judgment, we can do the same thing with an LLM. Researchers are very excited because the fluent, concise, informed, and perhaps even grounded responses from the LLM feel like interacting with a person, and so we guess they might have some of the same capabilities beyond producing fluent textual responses to prompts. In IR we are always complaining about the costs of human assessments, so perhaps this is solved. I would like to point out, although it is not the main thrust of this talk, that if the above is true, IR is solved and we don’t need to have research about it any more. The computer understands the document and the user information need to the degree that it can accurately predict if the document meets the need, and that is what IR systems are supposed to do. Scaling current LLM capabilities to where it can run on your wristwatch is just engineering. The actual thrust of this talk will be to review some of the history and literature on automatic evaluation methods. This is not automatic evaluation’s first rodeo, as they say. My arrival at NIST was accompanied by a SIGIR paper proposing that relevant documents could be picked using random sampling, and from that point the race was on. Along the way we have reinforced some things we already knew, like relevance feedback is good, and found some new things we did not know.

\subsubsection{\textbf{Keynote 2:} LLMs as Rankers, Raters, and Rewarders \\ \textit{by} Donald Metzler, Google DeepMind}
\textbf{Abstract.} In this talk, I will discuss recent advancements in the application of large language models (LLMs) to ranking, rating, and reward modeling, particularly in the context of information retrieval tasks. I will emphasize the fundamental similarities among these problems, highlighting that they essentially address the same underlying issue but through different approaches. Based on this observation, I propose several research questions that offer promising avenues for future exploration.

\subsection{Papers}
\label{sec:papers}
The workshop received 21 paper submissions, each of which was reviewed through EasyChair\footnote{\url{https://easychair.org/}} in a double-blind process by at least three reviewers from the list in Section \ref{sec:reviewers}. Reviewers rated papers as reject, weak reject, weak accept, or accept, with no option for a neutral (borderline) stance. Papers with mixed reviews were evaluated further by the organizers, who acted as meta-reviewers. We encouraged authors to include code and reproducibility efforts in their submissions.

All accepted papers are hosted on our website\footnote{\url{https://llm4eval.github.io/papers/}} in a non-archival format and were presented in a poster session. 7 of these papers were selected for presentation and publication in the CEUR-WS volume ``\textit{Proceedings of the 1st Workshop on Large Language Models for Evaluation in Information Retirveal (LLM4Eval)}''. Other 11 papers where accepted for presentation only. Additionally, the workshop received 5 already published works which, being on topic, were accepted for presentation to inspire group discussions. Although the submissions varied in perspectives, they all focused on evaluation topics. Below, we present the titles and authors of the accepted papers.

\subsubsection{Accepted Papers}
\label{sec:accepted-papers}

\begin{enumerate}
    \item \href{https://arxiv.org/abs/2302.11266}{One-Shot Labeling for Automatic Relevance Estimation} \citep{macavaney2023one} \\ \textit{Sean MacAvaney and Luca Soldaini}
    
    \item \href{https://dl.acm.org/doi/abs/10.1145/3539618.3591979}{Evaluating Cross-modal Generative Models Using Retrieval Task} \citep{bithel2023evaluating} \\ \textit{Shivangi Bithel and Srikanta Bedathur}
    
    \item \href{https://arxiv.org/abs/2404.04044}{A Comparison of Methods for Evaluating Generative IR} \citep{arabzadeh2024comparison} \\ \textit{Negar Arabzadeh and Charles L.~A.~Clarke}
    
    \item \href{https://www.arxiv.org/pdf/2408.01723}{A Novel Evaluation Framework for Image2Text Generation} \citep{huang2024novelevaluationframeworkimage2text} \\ \textit{Jia-Hong Huang, Hongyi Zhu, Yixian Shen, Stevan Rudinac, Alessio M.~Pacces and Evangelos Kanoulas}
    
    \item \href{https://arxiv.org/pdf/2407.13166}{Using LLMs to Investigate Correlations of Conversational Follow-up Queries with User Satisfaction} \citep{kim2024using} \\ \textit{Hyunwoo Kim, Yoonseo Choi, Taehyun Yang, Honggu Lee, Chaneon Park, Yongju Lee, Jin Young Kim and Juho Kim}
    
    \item \href{https://www.cs.unh.edu/~dietz/papers/farzi2024exampp.pdf}{EXAM++: LLM-based Answerability Metrics for IR Evaluation} \citep{farzi2024exam++} \\ \textit{Naghmeh Farzi and Laura Dietz}
    
    \item \href{https://arxiv.org/abs/2404.09980}{Context Does Matter: Implications for Crowdsourced Evaluation Labels in Task-Oriented Dialogue Systems} \citep{siro2024context} \\ \textit{Clemencia Siro, Mohammad Aliannejadi and Maarten de Rijke}
    
    \item \href{https://arxiv.org/abs/2405.00982}{On the Evaluation of Machine-Generated Reports} \citep{mayfield2024evaluation} \\ \textit{James Mayfield, Eugene Yang, Dawn Lawrie, Sean MacAvaney, Paul McNamee, Douglas W. Oard, Luca Soldaini, Ian Soboroff, Orion Weller, Efsun Kayi, Kate Sanders, Marc Mason and Noah Hibbler}
    
    \item \href{https://www.arxiv.org/pdf/2408.01363}{Toward Automatic Relevance Judgment using Vision–Language Models for Image–Text Retrieval Evaluation} \citep{yang2024automaticrelevancejudgmentusing} \\ \textit{Jheng-Hong Yang and Jimmy Lin}
    
    \item \href{https://arxiv.org/abs/2407.02464}{Reliable Confidence Intervals for Information Retrieval Evaluation Using Generative A.I.} \citep{oosterhuis2024reliable} \\ \textit{Harrie Oosterhuis, Rolf Jagerman, Zhen Qin, Xuanhui Wang and Michael Bendersky}
    
    \item \href{https://www.arxiv.org/abs/2405.06093}{Selective Fine-tuning on LLM-labeled Data May Reduce Reliance on Human Annotation: A Case Study Using Schedule-of-Event Table Detection} \citep{kumar2024selective} \\ \textit{Bhawesh Kumar, Jonathan Amar, Eric Yang, Nan Li and Yugang Jia}
    
    \item \href{https://arxiv.org/abs/2403.15246}{FollowIR: Evaluating and Teaching Information Retrieval Models to Follow Instructions} \citep{weller2024followir} \\ \textit{Orion Weller, Benjamin Chang, Sean MacAvaney, Kyle Lo, Arman Cohan, Benjamin Van Durme, Dawn Lawrie and Luca Soldaini}
    
    \item \href{https://arxiv.org/abs/2406.03339}{The Challenges of Evaluating LLM Applications: An Analysis of Automated, Human, and LLM-Based Approaches} \citep{abeysinghe2024challenges} \\ \textit{Bhashithe Abeysinghe and Ruhan Circi}
    
    \item \href{https://arxiv.org/pdf/2405.05600}{Can We Use Large Language Models to Fill Relevance Judgment Holes?} \citep{abbasiantaeb2024can} \\ \textit{Zahra Abbasiantaeb, Chuan Meng, Leif Azzopardi and Mohammad Aliannejadi}
    
    \item \href{https://arxiv.org/abs/2404.01012}{Query Performance Prediction using Relevance Judgments Generated by Large Language Model} \citep{meng2024query} \\ \textit{Chuan Meng, Negar Arabzadeh, Arian Askari, Mohammad Aliannejadi and Maarten de Rijke}
    
    \item \href{https://arxiv.org/abs/2406.00247}{Large Language Models for Relevance Judgment in Product Search} \citep{mehrdad2024large} \\ \textit{Navid Mehrdad, Hrushikesh Mohapatra, Mossaab Bagdouri, Prijith Chandran, Alessandro Magnani, Xunfan Cai, Ajit Puthenputhussery, Sachin Yadav, Tony Lee, Chengxiang Zhai and Ciya Liao}
    
    \item \href{https://arxiv.org/abs/2406.06458}{Evaluating the Retrieval Component in LLM-Based Question Answering Systems} \citep{alinejad2024evaluating} \\ \textit{Ashkan Alinejad, Krtin Kumar and Ali Vahdat}
    
    \item \href{https://arxiv.org/abs/2406.07299v1}{Exploring Large Language Models for Relevance Judgments in Tetun} \citep{de2024exploring} \\ \textit{Gabriel de Jesus and Sérgio Nunes}
    
    \item \href{https://arxiv.org/pdf/2406.15264}{A Comparative Analysis of Faithfulness Metrics and Humans in Citation Evaluation} \citep{zhang2024finegrainedcitationevaluationgenerated} \\ \textit{Weijia Zhang, Mohammad Aliannejadi, Jiahuan Pei, Yifei Yuan, Jia-Hong Huang and Evangelos Kanoulas}
    
    \item \href{https://arxiv.org/abs/2406.14783}{Evaluating RAG-Fusion with RAGElo: an Automated Elo-based Framework} \citep{rackauckas2024evaluating} \\ \textit{Zackary Rackauckas, Arthur Câmara and Jakub Zavrel}
    
    \item \href{#}{Enhancing Demographic Diversity in Test Collections Using LLMs}\\ \textit{Marwah Alaofi, Nicola Ferro, Paul Thomas, Falk Scholer and Mark Sanderson}
    
    \item \href{#}{GPT-4 Relevance Labelling can be Fooled by Query Keyword Stuffing}\\ \textit{Marwah Alaofi, Paul Thomas, Falk Scholer and Mark Sanderson}
\end{enumerate}

\subsection{Poster Session}
In light of the acceptance of 23 papers, we organized a dynamic poster session to facilitate the dissemination of their findings. All the presentations where held in person. Additionally, we integrated our poster session with two other SIGIR workshops, namely IR-RAG\footnote{\url{https://coda.io/@rstless-group/ir-rag-sigir24}} and ReNeuIR\footnote{\url{https://reneuir.org/}}, fostering a collaborative environment for sharing insights.

\subsection{Panel Discussion}
\label{sec:panel-discussion}
The workshop included a panel discussion on relevant topics raised by the audience concerning the use of the LLMs for evaluation. The invited panellists were Charlie Clarke (University of Waterloo), Laura Dietz (University of New Hampshire), Michael D. Ekstrand (Drexel University), and Ian Soboroff (National Institute of Standards and Technology (NIST)). The moderator was Bhaskar Mitra (Microsoft Research). 

\paragraph{Evaluation Validity.}
A large part of the discussion focused on the validity of the evaluation using LLMs. 
One thing that we should address if we envision the use of LLMs as assessors is the circularity of the evaluation. While it is true that, based on the TREC paradigm, some specific IR models are used to construct the pool of documents to be annotated, it was also demonstrated that the TREC-style evaluation does not introduce any form of bias towards such models. On the contrary, if we were to use an LLM both as an assessor and as a ranker, we could expect such a model to be favoured over other evaluated models. This might also impair the development of new LLMs if we were to evaluate them on judgements constructed using a simpler LLM. If we assume a similar evaluation protocol, an LLM would be considered perfect if it behaves exactly as the LLM used for the annotations, which might be worse and therefore suboptimal.

\paragraph{Intrinsic Randomness of the LLMs.}
A second element discussed during the panel concerned the intrinsic randomness of these models. Indeed, some operations that are becoming more and more common when operating with an LLM, such as prompt engineering or parameter tuning, induce randomness in the generation: it is impossible to know beforehand what the output will be, given a certain prompt. 
To address this limitation, one of the proposed solutions involved the development of repositories of baseline prompts for a series of tasks that should be as minimal as possible. In this regard, there was a consensus on the fact that some ``tricks'' that are known to work in practice should be avoided to build a solid evaluation strategy. Examples of such tricks involve using special characters, ``threatening'' or ``flattering'' sentences and other word sequences that might work in practice in specific use cases but for which we are not able to devise a mathematical model describing why and how they work. 
To address the randomness intrinsic to the LLMs, a point raised by the audience concerned the possibility of exploiting it to build distributions of answers. For example, it is possible to envision a scenario in which, instead of interrogating a single LLM with a single prompt once, we could interrogate multiple models multiple times using multiple prompts, to construct a distribution of probability over the answers/relevance judgements that can be used to summarize the LLMs' opinion on the topic. A downside of this approach, as highlighted by the panellists, is the consumption of the LLMs. Indeed, LLMs are not only expensive from an economic perspective, but they have also an environmental impact. This side should be taken into consideration when using these models, especially if we consider resource-intensive procedures as the sampling.

\paragraph{Replicability and Reproducibility.}
Another important issue raised during the discussion concerned the replicability of the experiments that involve LLMs as assessors. The community should agree on policies and guidelines concerning proprietary models that cannot be reimplemented and replicated autonomously by the research community. We should foresee, address, and prevent possible scenarios in which changes in a proprietary model impact the scientific conclusions and findings of the papers that rely on such a model for the evaluation and empirical validation of the hypotheses.
In this regard, the Diversity, Equity and Inclusion aspects should also be taken into consideration: often, proprietary models are subject to costs that might not be sustainable for research groups with fewer economic resources. In this sense, our future evaluation protocols based on LLMs should be applicable regardless of the resources available to the different research groups.
A counter-argument that was raised in this regard, concerns the Cranfield paradigm and TREC-style evaluation. Akin to LLM-based evaluation, during its initial stages, a part of the research community considered TREC-style evaluation to be expensive, hard to replicate, and not deterministic, due to the partial annotation of the topics. With the development and refinement of the protocols, as well as the increasing familiarity of the researchers with this type of evaluation, TREC-style evaluation has become the de facto standard procedure. It is possible to envision a similar path also for LLMs-based evaluation.

\paragraph{The Parallelism Between Human and LLMs Assessment.}
Finally, an open issue concerns the parallelism between human and LLM annotations. One observation that was made is that we are somehow used to feeding ``prompts'' to ``black-box operators''. Indeed, this is for example what happens with the assessor guidelines commonly used by both the research and industry communities. For humans, it is common to ``experience'' the act of searching: the annotation process in this sense can be described as a form of generalization of the act of finding relevant information, which is indeed something the assessors have experienced in their lives. This is certainly not the case for the current LLMs who do not have empirical experience in the real world and therefore are not capable of generalizing something they cannot have experienced.
\section{\texttt{LLMJudge} Challenge}
\label{sec:llmjudge}
The goal of the challenge was to attract the attention of the community towards using LMs for evaluation and to release datasets that could later be used to enhance research in this area. The \texttt{LLMJudge} challenge reused the MS MARCO datasets \citep{nguyen2016ms} as the primary benchmark. The test queries were a mix of previous years' TREC 2023 Deep Learning Track (TREC DL '23) test sets, which were released along with a development set for fine-tuning or in-context learning purposes. Participants were given a set of \texttt{$\langle$query, document$\rangle$} pairs and were asked to generate a relevance label.

Participants needed to submit their exact prompt together with the predicted labels for the documents. When submitting prompts, participants were also able to indicate the exact LLM model and parameters they employed to generate the run, which could be used to reproduce it. By allowing participants to submit their prompts, we could further analyze how these prompts might work across a variety of different LLM models.

In order to evaluate the quality of the generated labels, we used Cohen's $\kappa$ to see the labeler's agreement with \texttt{LLMJudge} test data at query-document level and the Kendall's $\tau$ to check the labeler's agreement with \texttt{LLMJudge} test data on system ordering, i.e., the runs that submitted to TREC DL 2023. In total, we had 39 submissions (i.e., the 39 labelers) from 7 groups from National Institute of Standards and Technology (NIST), RMIT University, The University of Melbourne, University of New Hampshire, University of Waterloo, Included Health, and University of Amsterdam.

\begin{figure}[ht]
    \centering
    \includegraphics[width=0.5\linewidth]{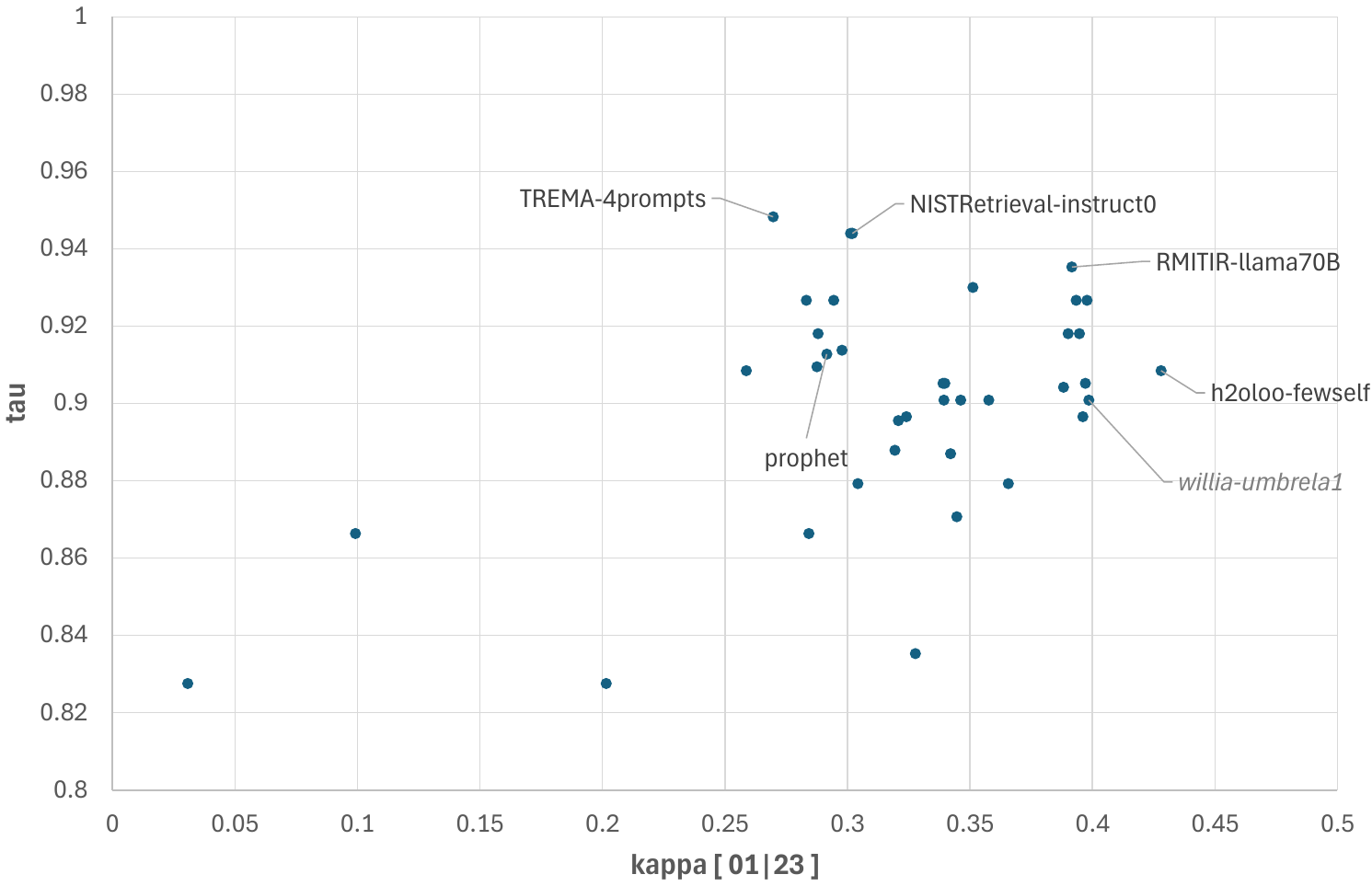}
    \caption{Scatter plot of Cohen's $\kappa$ and Kendall's $\tau$ for submitted labelers}
    \label{fig:llmjudge-plot}
\end{figure}

Figure \ref{fig:llmjudge-plot} shows the performance of submitted labelers on \texttt{LLMJudge} test set. The x-axis indicates Cohen's $\kappa$ while the y-axis shows the labeler's agreement on system ordering. It can be seen that labelers have a low variability for Kendall's $\tau$ but a larger for Cohen's $\kappa$. Most of the labelers are clustered in a narrow range of $\tau$ values, indicating that while they agree well on the ordering of systems, there is more variation in their inter-rater reliability as measured by Cohen's $\kappa$. This suggests that while the labelers tend to rank systems similarly, there is less consistency in their exact labeling, leading to variability in Cohen's $\kappa$ values.
\section{Conclusion}
\label{sec:conclusion}

The \workshopname 2024 workshop was designed as a platform to foster collaboration between academia and industry researchers from diverse backgrounds, united by a shared interest in the concept, development, and application of large language models for evaluation in information retrieval. This commitment to inclusivity is reflected in our workshop program, which features a panel comprising 4 researchers, a poster session showcasing 22 accepted papers, and a roundtable discussion. The immense potential of large language models in information retrieval and their subsequent applications in downstream services is widely recognized.
\section*{Acknowledgments}
We would like to thank ACM and SIGIR for hosting this workshop and extend our appreciation to the SIGIR 2024 workshop chairs, Zhuyun Dai, Hussein Suleman, and Andrew Trotman as well as our exceptional panellists, program committee members, paper authors, and participants. This research is supported by the Engineering and Physical Sciences Research Council [EP/S021566/1], the EPSRC Fellowship titled ``Task Based Information Retrieval'' [EP/P024289/1].

\bibliography{reference}
\end{document}